\documentclass{aa}

\usepackage{times}
\usepackage{hyperref}
\usepackage{graphics}

\thesaurus {05(08.16.5; 13.18.5; 13.25.5)}

\begin{document}

\title{Radio emission from RASS sources  south of  Taurus-Auriga}

\author {A. Magazz\`u\inst{1,2}\and G. Umana\inst{3} \and E.L. Mart{\'\i}n\inst{4}}

\institute {Centro Galileo Galilei, Apartado 565, E-38700 Santa Cruz de La Palma, Spain \and Osservatorio Astrofisico di Catania, Citt\`a Universitaria, I-95125 Catania, Italy \and Istituto di Radioastronomia del CNR, VLBI Station, CP 141, I-96017 Noto (SR), Italy \and University of California, 601 Campbell Hall, 94720 Berkeley, USA}

\offprints {A. Magazz\`u, address 1}
\mail{magazzu@tng.iac.es}

\date {Received  date /Accepted date}

\titlerunning {Radio emission south of Taurus-Auriga}
\authorrunning{A. Magazz\`u et al.}
\maketitle

\begin{abstract}

We present  a 8.4 GHz VLA survey of 50  optical counterparts to 46 ROSAT All-Sky survey (RASS) X-ray sour\-ces south of the Taurus-Auriga star-forming region. This survey detected 3 sources with a sensitivity limit of $\sim 0.12~\mbox{mJy}$. Merging our sample with other radio observations  of sources in and around Taurus-Auriga, we find scarce radio emission among RASS sources south of the Taurus-Auriga molecular clouds. Our data support the  evidence that these sources are older than weak-lined T Tauri stars, most of them probably being   close to the ZAMS.


\end{abstract}

\section{Introduction}
   
Recently,  the ROSAT All-Sky Survey (RASS) has shown the existence of a widely dispersed population of X-ray active stars  around nearby star-forming regions. This discovery has opened a lively debate on the nature of such objects.  After the first proposal that these sources are pre-main sequence (PMS) stars far from star-forming regions (Neuh\"auser et al.\ \cite{neuhausera}), some groups have  argued that these are young stars close to the main sequence, aged up to $10^8~\mbox{yr}$ (e.g.\ 
Brice\~no et al.\ \cite{briceno}). The study of the radio properties of the RASS sources can cast new light on this debate. 

Carkner et al.\ (\hyperlink{C97}{1997}, hereinafter \hyperlink{C97}{C97})  performed a 8.4 GHz VLA survey of 91 RASS sources, in the vicinity of the Taurus-Auriga star-forming cloud cores.  The optical counterparts of these sources had been classified as PMS objects  by Neuh\"auser et al.\ (\cite{neuhauserb}) and Wichmann et al.\ (\cite{wichmann}). Carkner et al.\ (\hyperlink{C97}{1997}) get a $3\sigma$ detection rate of 32\%, with a sensitivity limit of $\sim 0.15~\mbox{mJy}$.
They find no systematic difference between the sample of detected and undetected stars, as far as their spatial distribution is concerned. Instead, when considering the radio fluxes,  \hyperlink{C97}{C97} outline a significant difference of their sample from stellar groups like Pleiades ZAMS and active main sequence stars, and a similarity with weak-lined T Tauri stars (WTTS) previously detected at radio wavelengths.  In their conclusions they support the identification of most RASS sources around star-forming regions as genuine WTTS rather than ZAMS stars.

Independently from \hyperlink{C97}{C97}, we have conducted a 8.4 GHz VLA survey of 50 optical counterparts to 46 RASS sources south of Taurus-Auriga. According to the strength of the \ion{Li}{i} $\lambda 670.8~\mbox{nm}$ doublet in their spectrum, 30  of these objects had been classified by Magazz\`u et al.\  (\cite{magazzu}) as genuine PMS stars (100\% of their PMS stars), and the other 20  as possible PMS stars (87\% of their possible PMS stars). Twelve objects  resulted in common with \hyperlink{C97}{C97}. In this paper we present our observations, together with a new classification for the objects in our sample, and discuss our results  in the light of the debate on the nature of RASS sources around star-forming regions.

\section{Observations}

\label{obs}

The observations were carried out by using the VLA at 8.4 GHz with a total bandwidth of 100 MHz (two contiguous  50  MHz bands), in three different runs, namely 1997 February 18, February 20 and March 9. For all the observing runs the array was in B configuration. The sky was reported covered and partially  covered during the first two observational runs, i.e.\ on  February 18 and 20, respectively, being clear during the last run.

Each on-source  measurement consisted of a total of 14-15 minute integration time, preceded and followed by  2-3 minute observations of a  nearby phase calibrator. The  flux scale was fixed by daily observations of 3C48, whose flux density was assumed to be 3.28 Jy at 8.4 GHz. To avoid any spurious detection due to correlator offsets, for each pointing of our targets we shifted the phase centre by few arcsecs from the optical position.

The data were edited, calibrated and mapped using the Astronomical Image Processing System (AIPS), i.e. the standard software of the NRAO\@. To achieve the highest possible signal-to-noise ratio, the mapping process was performed by using the natural weighting and the dirty map was  CLEANed down as close as possible to the theoretical noise. We estimated the noise level in the maps by analyzing a map area reasonably far away from the phase centre and checked its consistency with the expected theoretical noise. Typical root-mean square (rms) of the noise level of the maps was of the order of 40 $\mu\mbox{Jy beam}^{-1}$, as can be seen in Table~\ref{sample}, where the rms is reported for each observation. This corresponds to a sensitivity limit ($3 \sigma$) of $\sim 0.12~\mbox{mJy}$.

\begin{table*}
\caption[]{Coordinates of the optical counterparts of the RASS sources, rms of our VLA maps and PMS classification (n.a.\ means that the used classification is not applicable, i.e.\ these stars are either PMS or ZAMS stars with spectral type F or G). A ``rs'' in the fourth column indicates radio sources detected in the map. Asterisks indicate stars in common with \hyperlink{C97}{C97}}

\label{sample}
\begin{flushleft}
\begin{tabular}{lrrlr}
\hline
RASS name   &   $\alpha~ (2000)$  &    $\delta~ (2000) $ &  rms & classification  \\
~      &                     &                    &  $\mu\mbox{Jy}$    \\
\hline                        
  \object{RX J0219.7$-$1026}$^*$  & 2 19 47.38 & $-$10 25 39.9      &   39     &PMS? \\
  \object{RX J0229.5$-$1224}  & 2 29 35.05 & $-$12 24 08.8              &    41     &n.a.\\
  \object{RX J0312.8$-$0414}NW& 3 12 50.43 & $-$04 14 08.1        &   46     &n.a.\\
 \object{RX J0312.8$-$0414}SE & 3 12 51\phantom{.00} & $-$04 14 19\phantom{.0}  &   46     &n.a.\\
  \object{RX J0324.4+0231}$^*$  & 3 24 25.22 &  02 31 01.2                &  36 rs  & PTTS \\
  \object{RX J0329.1+0118}$^*$   & 3 29 08.01 &  01 18 05.3              &  39       & n.a.\\
  \object{RX J0333.0+0354}$^*$  & 3 33 01.54 &  03 53 38.4               & 38        & PTTS \\
  \object{RX J0333.1+1036}$^*$   & 3 33 11.62 &  10 35 56.3              & 35        &PTTS\\
  \object{RX J0338.3+1020}  & 3 38 18.21 &  10 20 17.1                        & 38        &n.a.\\
  \object{RX J0339.6+0624}$^*$   & 3 39 40.57 &  06 24 43.1              &  40      &n.a.\\
  \object{RX J0343.6+1039}  & 3 43 40.51 &  10 39 14.0                        & 38 rs      &PMS? \\
  \object{RX J0344.8+0359}$^*$  & 3 44 53.15 &  03 59 30.9               & 37       &PMS? \\
  \object{RX J0347.2+0933}SW& 3 47 14.25 &  09 32 53.0                   &  36      &WTTS \\
 \object{RX J0347.2+0933}NE & 3 47 17\phantom{.00} &  09 33 08\phantom{.0}  &  36      & n.a.  \\
  \object{RX J0347.9+0616}  & 3 47 56.81 &  06 16 07.0                        &  39      &n.a.\\
  \object{RX J0348.5+0832}  & 3 48 31.42 &  08 31 37.5                        &  39     &n.a.\\
  \object{RX J0351.8+0413}  & 3 51 49.40 &  04 13 30.7                        & 37 rs  &n.a.\\
  \object{RX J0352.4+1223}  & 3 52 24.68 &  12 22 44.2                        &  36    &n.a.\\
  \object{RX J0354.1+0528}  & 3 54 06.58 &  05 27 23.7                        &  42  rs&n.a. \\
  \object{RX J0354.3+0535}$^*$   & 3 54 21.28 &  05 35 40.9             & 39      &n.a.\\
  \object{RX J0357.3+1258}  & 3 57 21.34 &  12 58 17.3                       & 44     &n.a.\\ 
  \object{RX J0358.1+0932}$^*$   & 3 58 12.62 &  09 32 21.6             & 37    &PMS? \\
  \object{RX J0400.1+0818}N & 4 00 09.37 &  08 18 19.3                     &  40    & WTTS\\
 \object{RX J0400.1+0818}S & 4 00 09\phantom{.00}  &  08 18 15\phantom{.0}   &  40   & PMS?  \\
  \object{RX J0404.4+0519}  & 4 04 28.48 &  05 18 43.7                       & 37     &PMS? \\
  \object{RX J0407.2+0113}N & 4 07 16.49 &  01 13 14.5                    &    39   &n.a.    \\
\object{RX J0407.2+0113}S& 4 07 16\phantom{.00}   &  01 13 12\phantom{.0}   &    39   &    PMS? \\
  \object{RX J0409.8+1209}  & 4 09 51.54 &  12 09 02.3                      &   39    &n.a. \\
  \object{RX J0410.6+0608}  & 4 10 39.66 &  06 08 38.9                      &   40     &PMS? \\
  \object{RX J0422.9+0141}  & 4 22 54.61 &  01 41 32.1                       &  33    &n.a. \\
  \object{RX J0423.5+0955}  & 4 23 30.28 &  09 54 30.0                       &  36 rs &PMS? \\
  \object{RX J0425.5+1210}  & 4 25 35.30 &  12 10 00.0                      &    33  &n.a.  \\
  \object{RX J0427.4+1039}  & 4 27 30.29 &  10 38 48.9                      &    37 &n.a. \\
  \object{RX J0427.5+0616}  & 4 27 32.07 &  06 15 52.1                      &    37 &n.a. \\
  \object{RX J0434.3+0226}  & 4 34 19.51 &  02 26 26.1                      &    36  &PTTS \\
  \object{RX J0442.5+0906}  & 4 42 31.94 &  09 06 01.7                       &   36  &n.a. \\
  \object{RX J0442.9+0400}  & 4 42 54.69 &  04 00 11.7                       &   38   &PMS? \\
  \object{RX J0444.4+0725}  & 4 44 27.16 &  07 24 59.8                      &    39 &PMS?  \\
  \object{RX J0444.7+0814}$^*$   & 4 44 45.44 &  08 13 47.6            &    37  &PMS? \\
  \object{RX J0445.2+0729}  & 4 45 13.18 &  07 29 17.7                      &    36 &n.a. \\
  \object{RX J0445.5+1207}  & 4 45 36.49 &  12 07 51.1                      &    38 & PTTS \\
  \object{RX J0450.0+0151}  & 4 50 04.68 &  01 50 43.1                      &    38  & WTTS \\
  \object{RX J0511.2+1031}  & 5 11 15.98 &  10 30 36.0                       &   39 &WTTS \\
  \object{RX J0511.9+1112}  & 5 12 00.30 &  11 12 19.8                       &   37&n.a. \\
  \object{RX J0512.0+1020}  & 5 12 03.21 &  10 20 06.8                       &   49&WTTS \\
  \object{RX J0516.3+1148}$^*$   & 5 16 21.52 &  11 47 47.3            &     37   &WTTS \\
  \object{RX J0528.5+1219}  & 5 28 35.19 &  12 19 04.0                       &    36 & WTTS\\
  \object{RX J0528.9+1046}  & 5 28 58.50 &  10 45 37.9                      &     37& WTTS\\
  \object{RX J0529.3+1210}  & 5 29 18.97 &  12 09 29.7                      &    39 & PTTS \\
  \object{RX J0531.8+1218}$^*$   & 5 31 47.77 &  12 18 08.1            &    40  & WTTS\\ 
\hline
\end{tabular}
\end{flushleft}
\end{table*}

Atmospheric phase fluctuations limit the sensitivity through decorrelation of visibilities. This effect, serious at higher frequencies, may be cause of failed detections of weak sources in extreme bad weather conditions. An inspection at the rms of calibrated phases in our data reveals, in the worst case, a rms of less than $20\degr$, that would cause an amplitude loss of few percents (Holdaway  \& Owen \cite {holdaway}).

 \section{Results}

In Table~\ref{sample} we list: name of the RASS source, with indication of the particular counterpart, in case of two counterparts to the same source ; coordinates of the optical counterpart;  the rms of the corresponding map (a ``rs'' indicates that one or more radio sources have been detected in the map), and the PMS classification of the optical counterpart, established according to the classification scheme proposed by Mart{\'\i}n (\cite{martin1}) and revised by Mart{\'\i}n \& Magazz\`u (\cite{martin}). 
These authors  indicate  that using higher resolution spectra and stricter criteria than in Neuh\"auser et al.\ (\cite{neuhauserb}) and Wichmann et al.\ (\cite{wichmann}), the number of genuine WTTS in the sample of Wichmann et al.\  can be reduced. In fact, observing 35 stars of Wichmann et al., Mart{\'\i}n \& Magazz\`u (\cite{martin}) confirm as WTTS less than  30\% of the objects, the rest resulting either  PTTS, or G and early-K stars for which the PMS status is dubious, or stars not showing lithium (7 objects). Note that Mart{\'\i}n \& Magazz\`u's (\cite{martin}) criteria translate into a different definition of WTTS than in  Neuh\"auser et al.\ (\cite{neuhauserb}), Wichmann et al.\ (\cite{wichmann}),  and \hyperlink{C97}{C97}, the main difference being that in Mart{\'\i}n \& Magazz\`u (\cite{martin}) WTTS and PTTS samples do not overlap.

Used acronyms are : CTTS= classical T Tauri stars; WTTS = weak-lined T Tauri stars; PTTS = post T Tauri stars (PMS stars with significant lithium depletion); PMS? = stars with lithium below the upper envelope defined by young open clusters. The  classification in Table~\ref{sample} refines and overrides the one given by Magazz\`u et al.\ (\cite{magazzu}) for these objects. Note that for G spectral types and earlier, both classifications, like any other based on the strength of the \ion{Li}{i} $\lambda 670.8~\mbox{nm}$ doublet,  are not applicable. In fact, stars hotter than about 5200 K do not suffer significant lithium depletion through the PMS evolution down to the main sequence. Stars in common with \hyperlink{C97}{C97}  are indicated by  asterisks. In our sample of 50 objects we have 9 WTTS, 6 PTTS, 12 PMS?, and 23 not classified star (indicated as ``n.a.'' in Table~\ref{sample}) .

In Table~\ref{new} we list the radio fluxes of the sources we found in the observed fields (level higher than $3 \sigma$), together with coordinates and offsets from the closest GSC stars. Carkner et al.\ (\hyperlink{C97}{1997}) distinguish reliable detections (at a level $\geq 4 \sigma$) and likely detections (at a level between 3 and $4 \sigma$; ``probable radio emitting'' stars). They also give  2\arcsec\ as maximum acceptable mismatch between optical position and radio source in order to consider the star as real optical counterpart. In this paper we will keep the same definitions, for better comparison with \hyperlink{C97}{C97} results, even if we would have been more conservative.  From Table~\ref{new} we see that two stars of our sample (GSC 66\_825, \object{RX J0343.6+1039}; GSC  86\_318, \object{RX J0434.3+0226}) can be identified with radio sources reliably detected, and one with a  source  likely  detected (GSC 91\_702, \object{RX J0442.9+0400}). The other radio sources are inside the uncertainty of the RASS source positions, so that radio emission can be still associated to X-ray emission, but not to the target stars. One may think  that in this case X-ray emission comes from the same source of radio  emission, probably extragalactic. If this is the case, we see that 4 out of 7 X-ray sources might not  be associated to their believed optical counterpart. A radio spectrum of these sources might help in clarifying this point. Finally, we note also that in the field of \object{RX J0324.4+0231}  a confusing source is present at 53 arcsecs from the target. This source has the same position as reported by \hyperlink{C97}{C97} and the same flux within the errors (cf.\ note on their Table 3).

\begin{table*}
\caption[]{Coordinates of radio emissions, their approximate offsets from GSC stars and radio fluxes}
\label{new}
\begin{flushleft}
\begin{tabular}{llrrll}
\hline
\multicolumn{2}{c}{Field} & $\alpha~ (2000)$  &    $\delta~ (2000) $ & offset & $f_{\rm r}$ \\
RASS &GSC     &                   &                      &    \arcsec    &  mJy  \\
\hline
\object{RX J0324.4+0231} &60\_489  &   03 24 23.53 &  02 31 48.01   &    53   &  $1.9^{\mathrm{a}}$  \\
\object{RX J0343.6+1039} & 66\_825  &   03 43 40.49    &  10 39 13.75     &  0.4  & 0.83    \\
\object{RX J0351.8+0413}& 69\_719  &   03 51 48.01 &  04 12 54.81    &   41   &  0.3 \\
\object{RX J0354.1+0528} &72\_606  &   03 54  06.90  &  05 27 30.04   &   7.9  &    0.3 \\
		&	           &   03 54  08.38  &  05 27 31.99    &  28.2 &    0.2  \\
		 &	           &   03 54  08.38  &  05 27 30.16    &  27.7 &    0.2  \\
 		&	           &   03 54 06.46   &  05 27 12.82    &  11   &  0.2  \\
 \object{RX J0434.3+0226} & 86\_318  & 04 34 19.53 & 02 26 24.60          &  1.5   & 0.15 \\
 \object{RX J0423.5+0955}  &72\_1156 &   04 23 30.40 &  09 54 15.35    &   14.7 &   $0.4^{\mathrm{b}}$ \\
 \object{RX J0442.9+0400}  & 91\_702 & 04 42 54.69 &  04 00 11.42  & 0.3 & $0.12^{\mathrm{c}}$ \\
\hline
\end{tabular}
\begin{list}{}{}
\item[$^{\mathrm{a}}$] Observed by \hyperlink{C97}{C97} with flux 2.0 mJy
\item[$^{\mathrm{b}}$] extended?
\item[$^{\mathrm{c}}$] likely detection (between 3 and $4 \sigma$)
\end{list}
\end{flushleft}
\end{table*}

 Note that 3 out of 12 sources in common with \hyperlink{C97}{C97} have been detected by these authors and not detected by us. Two of them (\object{RX J0219.7$-$1026} and \object{RX J0333.0+0354})  were detected by \hyperlink{C97}{C97} at 3.0 and $3.4 \sigma$, while the third source (\object{RX J0344.8+0359}) was detected at $4.8 \sigma$. In our maps these sources  should have been detected at $4.6 \sigma$,  $4.5 \sigma$,  and  $6.5 \sigma$, respectively. An obvious explaination of their absence in our maps is variability, which would argue for a young age of these three objects. In fact, those few radio-bright WTTS which have been carefully monitored have shown to be highly variable (see e.g.\ Stine et al.\ \cite{stine}, Cohen \& Bieging \cite{cohen}). However, we note the quite low level above the noise of the first two sources, which makes doubtful their detection.

\section {Discussion}

The $3 \sigma$ radio detection rate in our sample of RASS sources is 6\%, which is pretty  lower than the value reported by \hyperlink{C97}{C97} (32\%).  Once we merge \hyperlink{C97}{C97}'s  and our sample in a unique sample, the radio detection rate of RASS sources in and around Taurus-Auriga lowers from 32\% to 23\%, if we consider all the sources in common as undetected, or to 25\% if the three sources detected by \hyperlink{C97}{C97} and not detected by us are considered as detections. Incidentally,  after merging,   the detection rate of RASS sources in Taurus-Auriga becomes closer to 21\%,  the rate   found by Chiang et al.\ (\cite{chiang}) in previously known WTTS inside the Taurus complex.

\begin{figure}
\resizebox{\hsize}{!}{\includegraphics{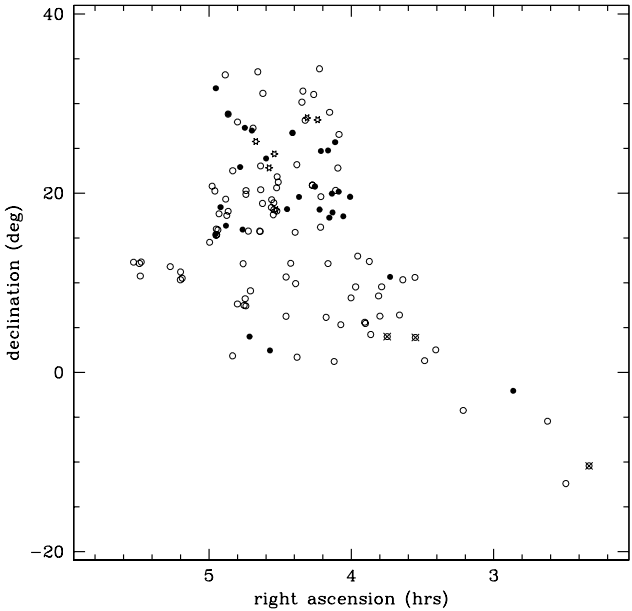}}
\caption[]{Position of the RASS sources observed in this work, together with those of \hyperlink{C97}{C97}. Radio detections are indicated by filled circles, non-detections by open circles. Crosses mark the three objects non detected by us and detected by \hyperlink{C97}{C97}. Star formation sites (from Gomez et al.\ \cite{gomez}) are also shown, indicated by starred symbols}
\label{map}
\end{figure}

In Fig.~\ref{map} we show the location of the RASS sources observed by us and by \hyperlink{C97}{C97}. At a first sight, we note a  difference in spatial distribution between radio detected and undetected RASS sources.  Namely, the radio detections drop  south of  about $15\degr$ of declination.This means that most of  them  are inside the Taurus-Auriga molecular clouds, as defined by the CO survey by Ungerechts \& Thaddeus (\cite{ungerechts}). In fact, north of $\delta = 15\degr$ the detection rate is 34\%, while south of that declination it is 8\% (13\% if we consider the RASS sources indicated with a cross in Fig.~\ref{map} as detected in radio).

O'Neal et al.\ (1990) suggested that T Tauri radio emission in Taurus occurs preferentially in the stars closer to the active cloud cores. Such suggestion was not confirmed by \hyperlink{C97}{C97}, who claim instead a uniform distribution of emitting and non-emitting sources. Our findings indicate that the uniform distribution claimed by \hyperlink{C97}{C97} may hold only inside the Taurus-Auriga molecular clouds, while, when extending the radio survey outside these clouds,  a clear lack of radio emission among RASS sources in the region south of Taurus-Auriga is found.

\begin{figure}
\resizebox{\hsize}{!}{\includegraphics{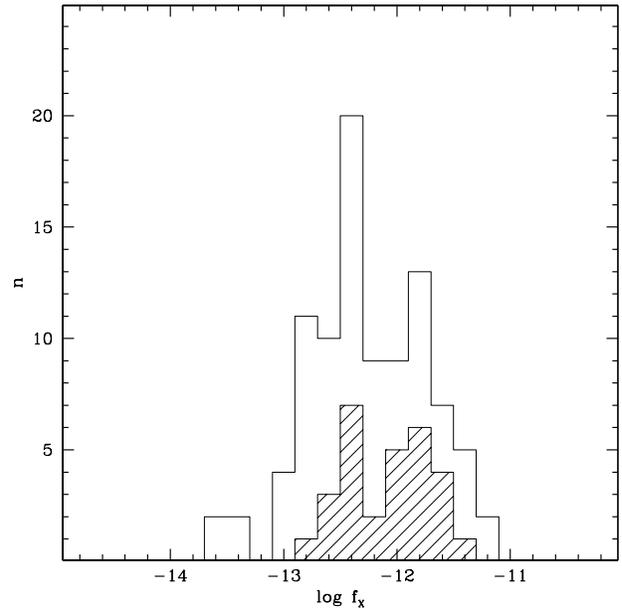}}
\caption[]{Distribution of X-ray fluxes of objects  undetected and detected (shaded) in radio. Units of flux are $\mbox {erg s}^{-1}~\mbox {cm}^{-2}$}
\label{histo}
\end{figure}

One may think that undetected radio sources are such because they are located at higher distances. In order to assess if this is the case, we plot in Fig.~\ref{histo} the distribution of X-ray fluxes of undetected and detected sources, from Magazz\`u et al.\ (\cite{magazzu}) and from X luminosities in \hyperlink{C97}{C97} (assuming $d = 140~\mbox{pc}$). We see that the two distributions are almost overlapping, suggesting that there is no significant difference in the average  distance of the two subsamples. As discussed in Section~\ref{obs}, we exclude that our non-detections are due to flaws in data acquisition or reduction. Variability could have also affected our results, but it is not reasonable to think that so many sources have varied downward, vanishing below the detection limit, at the time of our observations. Therefore, the only explanation for having only few detections in our sample seems to be the intrinsic low level or radio emission of our sources.

The $3 \sigma$ upper limits ($3 \times$ rms in Table~\ref{sample}), translated to luminosity upper limits, tell that the level of radio emission in most of the stars of  the southern sample might be similar to  what is observed in  coronally active K, G, and F main sequence stars (G\"udel \cite{gudel} and G\"udel et al.\ \cite{gudel1}), or -- at most -- in Pleiades stars (Bastian et al.\ \cite {bastian}, Lim \& White \cite{lim}). This argues for an older age of the southern sample with respect to RASS sources inside Taurus-Auriga. In fact, \hyperlink{C97}{C97} find that, as far as  the radio properties are concerned, their sample (mainly composed of stars inside Taurus-Auriga) is  very similar to other samples of WTTS in the literature, and clearly different  from  active main sequence stars and Pleiades ZAMS. Our estimate of an older age for our sample could be supported by Frink et al.'s (\cite{frink}) results that the stars in the southern part of Taurus-Auriga are kinematically different from the stars in the central region. All the attempts made by these authors to kinematically relate the optical counterparts of the RASS sources south of the Taurus-Auriga cloud to the central part of this associations have failed. They conclude that two different star formation scenarios are to be invoked for these two areas.

These indications on the age of the RASS sources south of Taurus-Auriga  have to be compared with previous age determinations, mainly based on the equivalent width of the \ion{Li}{i} $\lambda 670.8~\mbox{nm}$ doublet. According to Magazz\`u et al.\  (\cite{magazzu}), 61\% of the stars observed here are PMS stars, the rest being stars in which lithium has been detected at a level similar to Pleiads of the same spectral type, and then presumably older than PMS stars. Results of a more conservative classification, performed according to the scheme proposed by Mart{\'\i}n \& Magazz\`u (\cite{martin}), are presented in Table~\ref{sample}. From this table we see that the number of WTTS is  18\% of the sample\footnote{These are lower limits, since no classification has been given to the F and G stars, i.e.\ to 46\% of the sample.}. If we consider  PTTS as a subsample of PMS stars, we have that PMS stars constitute 56\% of the sample of K and M stars (and 30\% of the total sample). A determination of age of these southern objects has been performed also by Neuh\"auser et al.\ (\cite{neuhauser97}). According to their estimates, the age of 31\% of our sample  results to be lower than $3 \times 10^7~\mbox{yr}$ (PMS stars). The remaining stars were classified as ZAMS stars or could not be classified due to poor  signal-to-noise ratio or low resolution spectra. Therefore, we can conclude that spectroscopic classifications and radio observations agree in assigning to the stars of our sample an age on average older than WTTS in Taurus-Auriga, close to the main sequence in most of the cases. A different sample of 35 RASS sources (with only one source in common with our work) in an area $20\degr$ south of the Taurus-Auriga star-forming region has been studied by Zickgraf et al.\ (\cite{zickgraf}). They find 9 stars with age 10-30 Myr, estimating that this number of stars with such an age is in excess with respect to standard models of stellar galactic distribution.

\begin{table}
  \caption[]{Classification (from Mart\'\i n \& Magazz\`u \cite{martin}; n.a.\ means that the classification is not applicable) and radio flux (from \hyperlink{C97}{C97}) of some stars in the \hyperlink{C97}{C97} sample. Number is after Wichmann et al.\ (\cite{wichmann})}
\label{nature}
\begin{flushleft}
  \begin{tabular}{llrc}
\hline
Number & RASS name     & classification& radio flux \\
	& 	& & mJy \\
\hline
W2    &  \object{RX J0403.3+1725}   & WTTS   &    0.22 \\
W3    &  \object{RX J0405.1+2632}   & PMS?  &          \\
W7    &  \object{RX J0406.8+2541}   & WTTS   &  0.29       \\
W8    &  \object{RX J0407.8+1750}   & PMS?   &    0.18     \\
W9    &  \object{RX J0408.2+1956}   & dK  &     0.19    \\
W12   &  \object{RX J0409.8+2446}   & PTTS   &  0.57       \\
W13   &  \object{RX J0412.8+1937}   & PTTS   &           \\
W14   &  \object{RX J0412.8+2442}   & n.a.   &   0.35 \\
W15   &  \object{HD 285579}        & n.a.   &           \\
W16   &  \object{RX J0413.3+1810}   & dMe    &   0.29      \\
W20   &  \object{RX J0416.5+2053A}   & WTTS   &           \\
W21   &  \object{RX J0416.5+2053B}   & WTTS   &           \\
W22   &  \object{RX J0419.4+2808}   & n.a.  &           \\
W24   &  \object{RX J0420.8+3009}   & PTTS   &           \\
W25   &  \object{RX J0422.1+1934}   & WTTS   &  0.19       \\
W28   &  \object{BD +26 718}       & WTTS   &   1.97      \\
W29   &  \object{BD +26 718B}      & WTTS   &   0.33      \\
W33   &  \object{RX J0431.3+1800}   & CTTS   &           \\
W34   &  \object{RX J0431.4+2035}   & PTTS   &           \\
W35   &  \object{RX J0432.6+1809}   & WTTS   &           \\
W36   &  \object{RX J0432.7+1853}   & PMS?   &           \\
W37   &  \object{RX J0432.8+1735}   & WTTS   &           \\
W39   &  \object{RX J0433.7+1823}   & n.a.    &           \\
W42   &  \object{RX J0437.4+1851A}   & PTTS   &           \\
W45   &  \object{RX J0438.2+2302}  & PTTS   &           \\
W49   &  \object{RX J0441.4+2715}   & n.a.   &           \\
W53   &  \object{RX J0444.4+1952}  & dMe    &           \\
W56   &  \object{RX J0446.8+2255}   & dMe    &    0.45     \\
W59   &  \object{RX J0451.8+1758}   & WTTS   &           \\
W60   &  \object{RX J0451.9+2849A}   & PMS?   &   0.18  \\
W61   &  \object{RX J0451.9+2849B}   & PMS?   &           \\
W69   &  \object{RX J0456.7+1521}   & dMe    &           \\
W72   &  \object{RX J0457.0+3142}   & dK   &     0.17    \\
W75   &  \object{RX J0458.7+2046}   & PTTS   &           \\
\hline
\end{tabular}
\end{flushleft}
\end{table}

It is worthwhile, at this point, to look at radio emission of \hyperlink{C97}{C97} stars in the light of the classification by Mart{\'\i}n \& Magazz\`u (\cite{martin}), shown in Table~\ref{nature} for 34 out of 91 \hyperlink{C97}{C97} stars. The radio flux is also shown, if detected at a level higher than $3 \sigma$.  We see that the detection rate among PMS stars (WTTS+CTTS+PTTS) is 0.33, similar to (and even lower than) the detection rate  (0.44) in the group of remaining objects. Moreover, radio fluxes in the two groups are quite similar. Therefore, there seems to be no significant difference between radio emission in PMS stars of Taurus-Auriga and other radio-emitting RASS sources found in the same region. It is interesting to note that for WTTS+CTTS the detection rate is 0.45, while for PTTS this rate lowers to 0.14.

\section {Conclusions}

We have observed with VLA 50 optical counterparts to 46 RASS sources south of Taurus-Auriga. Merging our observations with those of Carkner et al.\ (\hyperlink{C97}{1997}) we obtain the following results:

\begin{itemize}

\item although spatial distribution of radio sources may be uniform among RASS sources  inside the Taurus-Auriga molecular clouds, RASS sources south of  the clouds show scarce radio emission;

\item  there is evidence that our RASS sources south of Taurus-Auriga are older than WTTS and sources inside the clouds.

\end{itemize}

Moreover, there seems to be no significant difference between radio emission in PMS stars of Taurus-Auruga and other radio-emitting RASS sources found in the same region. More radio observations of RASS sources in star-forming regions, together with accurate  classification of the optical counterparts will be essential in order to improve our understanding about the age, status, and origin of these objects.

\begin {acknowledgements}

We thank our referee, R.~Neuh\"auser, for a detailed report, which helped to improve this paper.
 
\end {acknowledgements}


\begin{thebibliography}{}

\bibitem[1988]{bastian} Bastian, T. S., Dulk, G. A., Slee, O. B. 1988, AJ 95, 794
\bibitem[1997]{briceno} Brice\~no, C., Hartmann, L. W., Stauffer, J.R., Gagn\`e, M., Stern R.A., Caillault, J.-P. 1997, AJ 113, 740
\bibitem{}\hypertarget{C97}{Carkner}, L., Mamajek, E., Feigelson, E., Neuh\"auser, R., Wichmann, R., Krautter, J. 1997, ApJ 490, 735 (C97)
\bibitem[1996]{chiang} Chiang, E., Phillips, R. B., Londsale, C. J. 1996, AJ 111, 355
\bibitem[1986]{cohen} Cohen, M., Bieging, J. H. 1986, AJ 92, 1396
\bibitem[1997]{frink} Frink, S., R\"oser, S., Neuh\"auser, R., Sterzik, M. F. 1997, A\&A 325, 613
\bibitem[1993]{gomez} Gomez, M., Hartmann, L., Kenyon, S. J., Hewett, R. 1993, AJ 105, 192
\bibitem[1992]{gudel} G\"udel, M., 1992, A\&A 264, L31
\bibitem[1995]{gudel1} G\"udel, M., Schmitt, J. H. M. M., Benz, A. O. 1995, A\&A 302, 775
\bibitem[1995]{holdaway} Holdaway,  M.,  Owen, F. 1995,  MMA Memo, 136
\bibitem[1995]{lim} Lim, J., White, S. M. 1995, ApJ 453, 207
\bibitem [1997]{magazzu} Magazz\`u, A., Mart{\'\i}n, E.L., Sterzik, M. F., Neuh\"auser, R., Covino, E., Alcal\'a, J. M. 1997, A\&AS 124, 449
\bibitem[1997]{martin1} Mart{\'\i}n, E. L. 1997, A\&A 321, 492
\bibitem[1999]{martin} Mart{\'\i}n, E. L., \& Magazz\`u, A. 1999, A\&A 342, 173
\bibitem[1995a]{neuhausera} Neuh\"auser, R., Sterzik, M. F.,  Schmitt, J. H. M. M., Wichmann, R., Krautter, J. 1995a, A\&A, 295, L5
\bibitem[1995b]{neuhauserb} Neuh\"auser, R., Sterzik, M. F., Torres, G., Mart{\'\i}n, E. L. 1995b, A\&A 229, L13
\bibitem[1997]{neuhauser97} Neuh\"auser, R., Torres, G.,  Sterzik, M. F., Randich, S. 1997, A\&A 325, 647
\bibitem [1990]{oneal}O'Neal, D., Feigelson, E. D., Mathieu, R. D., Myers, P. C. 1990, AJ 100, 1610
\bibitem[1988]{stine} Stine, P. C., Feigelson, E. D., Andr\'e, P., Montmerle, T. 1988, AJ 96, 1394
\bibitem [1987]{ungerechts} Ungerechts H., Thaddeus, P.  1987, ApJS 63, 645
\bibitem[1996]{wichmann} Wichmann, R., Krautter, J., Schmitt, J.H.M.M., et al.\ 1996, A\&A 312, 439
\bibitem[1998]{zickgraf} Zickgraf, F.-J., Alcal\'a, J. M., Krautter, J., Sterzik, M. F., Appenzeller, I., Motch, C., Pakull, M.W. 1998, A\&A 339, 457
\end{thebibliography}
\end{document}